\documentclass[english]{IEEEtran}
 \pdfoutput=1
\usepackage[T1]{fontenc}
\usepackage[latin9]{inputenc}
\usepackage{array}
\usepackage{units}
\usepackage{multirow}
\usepackage{amsmath}
\usepackage{amsthm}
\usepackage{amssymb}
\usepackage{stackrel}
\usepackage{graphicx}

\makeatletter

\providecommand{\tabularnewline}{\\}

\theoremstyle{definition}
\newtheorem{assumption}{Assumption}
\theoremstyle{definition}
\newtheorem{example}{\protect\examplename}
\theoremstyle{remark}
\newtheorem{rem}{\protect\remarkname}
\theoremstyle{plain}
\newtheorem{thm}{\protect\theoremname}

\usepackage{cite}
\usepackage{multirow}

\makeatother

\usepackage{babel}
\providecommand{\examplename}{Example}
\providecommand{\remarkname}{Remark}
\providecommand{\theoremname}{Theorem}

\begin{document}
\title{Lyapunov-Based Dropout Deep Neural Network (Lb-DDNN) Controller}
\author{Saiedeh Akbari$^{*}$, Emily J. Griffis$^{*}$, Omkar Sudhir Patil$^{*}$,
Warren E. Dixon$^{*}$\thanks{$^{*}$Saiedeh Akbari, Emily J. Griffis, Omkar Sudhir Patil, and Warren
E. Dixon are with the Department of Mechanical and Aerospace Engineering,
University of Florida, Gainesville, FL, 32611-6250 USA. Email: \{akbaris,
emilygriffis00, patilomkarsudhir, wdixon\}@ufl.edu.}\thanks{This research is supported in part by AFRL grant FA8651-21- F-1027,
AFRL grant FA8651-21-F-1025, AFOSR grant FA9550-19- 1-0169, and Office
of Naval Research grant N00014-21-1-2481. Any opinions, findings,
and conclusions or recommendations expressed in this material are
those of the author(s) and do not necessarily reflect the views of
the sponsoring agencies.}}
\maketitle
\begin{abstract}
Deep neural network (DNN)-based adaptive controllers can be used to
compensate for unstructured uncertainties in nonlinear dynamic systems.
However, DNNs are also very susceptible to overfitting and co-adaptation.
Dropout regularization is an approach where nodes are randomly dropped
during training to alleviate issues such as overfitting and co-adaptation.
In this paper, a dropout DNN-based adaptive controller is developed.
The developed dropout technique allows the deactivation of weights
that are stochastically selected for each individual layer within
the DNN. Simultaneously, a Lyapunov-based real-time weight adaptation
law is introduced to update the weights of all layers of the DNN for
online unsupervised learning. A non-smooth Lyapunov-based stability
analysis is performed to ensure asymptotic convergence of the tracking
error. Simulation results of the developed dropout DNN-based adaptive
controller indicate a $\boldsymbol{38.32\%}$ improvement in the tracking
error, a $\boldsymbol{53.67\%}$ improvement in the function approximation
error, and $\boldsymbol{50.44\%}$ lower control effort when compared
to a baseline adaptive DNN-based controller without dropout regularization.
\end{abstract}

\begin{IEEEkeywords}
Deep neural network, dropout, adaptive control, Lyapunov methods,
nonlinear control systems.
\global\long\def\SS{\mathbb{S}}%
\global\long\def\RR{\mathbb{R}}%
\global\long\def\nz{\left\Vert z\right\Vert }%
\global\long\def\ne{\left\Vert e\right\Vert }%
\global\long\def\nt{\left\Vert \widetilde{\theta}\right\Vert }%
\global\long\def\nzz{\left\Vert z\right\Vert ^{2}}%
\global\long\def\nee{\left\Vert e\right\Vert ^{2}}%
\global\long\def\ntt{\left\Vert \widetilde{\theta}\right\Vert ^{2}}%
\global\long\def\tq{\triangleq}%
\global\long\def\Linf{\mathcal{L}_{\infty}}%
\global\long\def\yy{\mathcal{Y}}%
\global\long\def\uu{\mathcal{U}}%
\global\long\def\grad{\mathcal{\nabla_{\widetilde{\theta}}}}%
\global\long\def\tvec{\mathcal{\text{vec}}}%
\end{IEEEkeywords}

\section{Introduction}

Empirical evidence indicates that deep neural networks (DNNs) can
provide better function approximation than single layer neural networks
\cite{Liang.Srikant2016}. Traditionally, DNN-based controllers are
trained using offline training methods based on prior collected datasets.
\cite[Section 6.6]{Brunton.Kutz2019}. Recent developments in \cite{Patil.Le.ea.2022,Sun.Greene.ea2021,Patil.Le.ea2022,Joshi.Chowdhary2019,Griffis.Patil.ea23_2}
use Lyapunov-based methods to develop unsupervised online learning
for all weights of a deep neural network (i.e., Lb-DNNs). 

Unfortunately, both offline and Lb-DNNs can exhibit significantly
degraded performance due to data overfitting. Another challenge that
decreases generalization and performance of the trained DNN is co-adaptation,
where multiple neurons, or even entire layers, become overly reliant
on each other during the training process\cite{Charu2018}. One effective
approach to address these issues is through dropout regularization.
Dropout was originally introduced by G. Hinton in \cite{Hinton.Srivastava.ea2012}
to prevent co-adaptation of feature detectors and improve the generalization
performance of DNNs. Dropout regularization involves stochastically
dropping out neurons during training, which helps prevent over-fitting,
enhances the overall function approximation performance, and efficiently
allocates computational resources while updating the network's weights
\cite{Le.Greene.ea2021,Srivastava.Hinton.ea2014}. By setting the
activation of certain individual weights to zero, dropout induces
sparse representation in the network which reduces co-dependency in
neurons. Moreover, dropouts can be viewed as training an ensemble
of multiple DNNs with smaller width that are trained independently.
Independence in the training has a regularizing effect and provides
better generalization to new information. This intuitive reasoning
is also applicable for using dropout in DNN-based adaptive control,
since dropouts mitigate co-adaptation by reducing the number of weights
influencing an adaptation law. 

Although dropout regularization has been used for offline training
of DNNs in results such as \cite{Ba.Frey2013,Dahl.Sainath.ea2013},
its application has been limited in real-time adaptive control settings.
In \cite{Le.Greene.ea2021}, the dropout method is employed on a DNN
to improve the training performance of inner layers in pseudo real-time,
and through simulations, the study demonstrates the improved performance
of a DNN-based adaptive controller with dropout. However, the pseudo
real-time adaptation laws in \cite{Le.Greene.ea2021} are not stability-driven
but are rather based on a modular design where the stability analysis
is primarily facilitated using robust control techniques.

This paper introduces a novel dropout technique aimed at enhancing
the function approximation performance of a DNN-based adaptive controller
that updates the weights of all layers of the DNN using the Lyapunov-based
update law in \cite{Patil.Le.ea.2022} (i.e., a Lyapunov-based Dropout
Deep Neural Network (Lb-DDNN)). The proposed technique involves the
selective inactivation, or dropout, of weights associated with randomly
selected neurons within each DNN layer. To incorporate dropout regularization,
a new recursive DNN representation and stability-driven weight adaptation
laws are constructed by considering the effect of randomization matrices
on the closed-loop error system. Through a non-smooth Lyapunov-based
stability analysis, the designed controller is guaranteed to stabilize
the system in the sense that the tracking error asymptotically converges
to zero. Simulation experiments are performed to compare the Lb-DDNN
adaptive controller with the baseline adaptive DNN controller developed
in \cite{Patil.Le.ea.2022}. The simulation results show a $35.56\%$
improvement in the tracking error, a $49.94\%$ improvement in the
function approximation error, and $48.56\%$ lower control effort
in the proposed controller when compared to the baseline controller.

\section{Problem Formulation\label{sec: problem formulation}}

\subsection{Notation\label{subsec:Notation}}

The space of essentially bounded Lebesgue measurable functions is
denoted by $\mathcal{L}_{\infty}$. Given two functions $f:A\to B$
and $g:B\to C$, where $A$, $B$, and $C$ are sets, the composition
of $f$ and $g$, denoted as $g\circ f$, is a new function $h:A\to C$
defined as $h\tq g\circ f=g\left(f\left(x\right)\right)$, for all
$x\in A$. Let $\mathbf{0}_{m\times n}$ denote a zero matrix with
the dimension of $m\times n$. Let $I_{n\times n}$ denote an identity
matrix with the dimension of $n$. For matrices $A\in\RR^{m\times n}$
and $B\in\RR^{p\times q}$, the Kronecker product is denoted as $A\otimes B$.
Given a matrix $A\triangleq\left[a_{i,j}\right]\in\mathbb{R}^{n\times m}$,
where $a_{i,j}$ denotes the element in the $i^{th}$ row and $j^{th}$
column of $A$, the vectorization operator is defined as $\tvec\left(A\right)\triangleq\left[a_{1,1},\ldots,a_{1,m},\ldots,a_{n,1},\ldots,a_{n,m}\right]^{\top}\in\mathbb{R}^{nm}$.
From \cite[Proposition 7.1.9]{Bernstein2009} and given matrices $A\in\mathbb{R}^{p\times a}$,
$B\in\mathbb{R}^{a\times r}$, and $C\in\mathbb{R}^{r\times s}$,
the vectorization operator satisfies the property $\tvec\left(ABC\right)=\left(C^{\top}\otimes A\right)\tvec\left(B\right).$
Differentiating $\tvec\left(ABC\right)$ on both sides with respect
to $\tvec\left(B\right)$ yields the property $\frac{\partial}{\partial\tvec\left(B\right)}\tvec\left(ABC\right)=C^{\top}\otimes A.$
The right-to-left matrix product operator is represented by $\stackrel{\curvearrowleft}{\prod}$,
i.e., $\stackrel{\curvearrowleft}{\stackrel[p=1]{m}{\prod}}A_{p}=A_{m}\ldots A_{2}A_{1}$,
and $\stackrel{\curvearrowleft}{\stackrel[p=a]{m}{\prod}}A_{p}=1$
if $a>m$. The Filippov set-valued map defined in \cite[Equation 2b]{Paden1987}
is denoted by $\text{K}\left[\cdot\right]$. The notation $\overset{\text{a.a.t.}}{\left(\cdot\right)}$
denotes that the relation $\left(\cdot\right)$ holds for almost all
time (a.a.t.). Consider a Lebesgue measurable and locally essentially
bounded function $h:\mathbb{R}^{n}\times\mathbb{R}_{\geq0}\to\mathbb{R}^{n}$.
Then, the function $y:\mathcal{I}\to\mathbb{R}^{n}$ is called a Filippov
solution of $\dot{y}=h\left(y,t\right)$ on the interval $\mathcal{I}\subseteq\mathbb{R}_{\geq0}$
if $y$ is absolutely continuous on $\mathcal{I}$ and $\dot{y}\overset{\text{a.a.t.}}{\in}\text{K}\left[h\right]\left(y,t\right)$.
Given some functions $f$ and $g$, the notation $f\left(y\right)=\mathcal{O}^{m}\left(g\left(y\right)\right)$
means that there exists some constants $M\in\mathbb{R}_{>0}$ and
$y_{0}\in\mathbb{R}$ such that $\left\Vert f(y)\right\Vert \leq M\left\Vert g(y)\right\Vert ^{m}$
for all $y\geq y_{0}$. The operator $\text{proj}\left(\cdot\right)$
denotes the projection operator defined in \cite[Appendix E, Eq. E.4]{Krstic1995}.

\subsection{Dynamic Model and Control Objective\label{subsec:Dynamic-Model}}

Consider a control-affine nonlinear system modeled as 
\begin{equation}
\dot{x}\left(t\right)=f\left(x\left(t\right)\right)+u\left(t\right),\label{eq: dynamics}
\end{equation}
where $t\in\RR_{\geq0}$, $x:\RR_{\geq0}\to\RR^{n}$, $f:\RR^{n}\to\RR^{n}$,
and $u:\RR_{\geq0}\to\RR^{n}$ denote continuous time, the state,
the unknown differentiable drift vector field, and the control input,
respectively. The control objective is to design a controller $u\left(t\right)$
such that the state tracks the desired trajectory $x_{d}$. To achieve
the control objective, an adaptive Lb-DNN architecture and a controller
are designed to learn the unknown drift vector field and to achieve
asymptotic convergence on the tracking error, respectively. To quantify
the control objective, the tracking error $e:\RR_{\geq0}\to\RR^{n}$,
is defined as
\begin{equation}
e\left(t\right)\tq x\left(t\right)-x_{d}\left(t\right),\label{eq: error}
\end{equation}
where $x_{d}:\RR_{\geq0}\to\RR^{n}$ denotes a continuously differentiable
desired trajectory. 
\begin{assumption}
\label{thm: boundedness of desired trajectory}The desired trajectory
is designed such that for all $t\in\RR_{\geq0}$, $x_{d}\left(t\right)\in\Omega$,
and $\dot{x}_{d}\in\Linf$, where $\Omega\subset\RR^{n}$ is a known
compact set. Hence, the desired trajectory can be bounded as $\left\Vert x_{d}\right\Vert \leq\overline{x_{d}}$,
where $\overline{x_{d}}\in\RR_{>0}$ is a known constant.
\end{assumption}

\section{Control Design}

\subsection{Deep Neural Network Architecture}

To estimate the unknown nonlinear drift vector field $f\left(x\right)$,
a Lb-DNN architecture is developed using dropout. Dropout randomly
omits neurons while training, which helps mitigate over-fitting and
co-adaptation, thus improving the overall performance and function
approximation capabilities of the DNN \cite{Le.Greene.ea2021,Srivastava.Hinton.ea2014}.
Leveraging the Lyapunov stability-driven weight adaptation laws developed
in \cite{Patil.Le.ea.2022}, the dropout DNN is designed such that
randomization matrices are used to incorporate dropout techniques
into the online, stability-driven weight adaptation. Through the randomization
matrices, weights associated with a batch of randomly selected neurons
are inactivated, i.e., dropped out, to reduce the interdependency
and excessive reliance on specific weights and neurons.

\begin{figure}
\centering{}\includegraphics[scale=0.35]{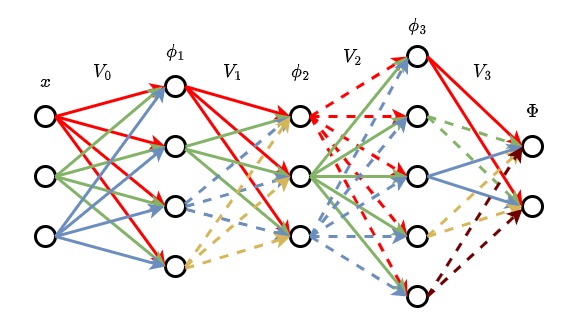}\caption{The structure of a DNN with three hidden layers, where the dashed
and solid lines respectively represent the randomly dropped out and
selected weights.\label{fig:RDNN arch.}}
\end{figure}

As shown in Figure \ref{fig:RDNN arch.}, let the dropout DNN architecture,
$\Phi:\RR^{n}\times\left\{ 0,1\right\} ^{\stackrel[m=0]{k}{\sum}L_{m}\times\stackrel[m=0]{k}{\sum}L_{m}}\times\RR^{\stackrel[j=0]{k}{\sum}L_{j}L_{j+1}}\to\RR^{L_{k+1}}$,
be defined as {\footnotesize{}
\begin{equation}
\Phi\left(x,R_{i},\theta\right)=\left(R_{i,k}V_{k}\right)^{\top}\phi_{k}\circ\cdots\circ\left(R_{i,1}V_{1}\right)^{\top}\phi_{1}\circ\left(R_{i,0}V_{0}\right)^{\top}x,\label{eq: DNN}
\end{equation}
}where $k\in\mathbb{Z}_{>0}$ denotes the number of the layers in
$\Phi\left(x,R_{i},\theta\right)$, and $\phi_{j}:\RR^{L_{j}}\to\RR^{L_{j}}$
denotes the vector of smooth activation functions in the $j^{\text{th}}$
layer, for all $j\in\left\{ 1,\cdots,k\right\} $.\footnote{Although $\phi_{j}$ is defined as a smooth function, the subsequent
analysis allows the inclusion of non-smooth activation functions by
using the switched systems analysis in \cite{Patil.Le.ea.2022}.} For $j\in\left\{ 0,\cdots,k\right\} $, $V_{j}\in\RR^{L_{j}\times L_{j+1}}$
and $L_{j}\in\mathbb{Z}_{>0}$ represent the weight matrix and the
number of nodes in the $j^{\text{th}}$ layer of $\Phi$, respectively.
For notation simplicity, the weights can be represented in a vector
$\theta\in\RR^{\stackrel[j=0]{k}{\sum}L_{j}L_{j+1}}$ as $\theta\tq\left[\tvec\left(V_{0}\right)^{\top},\tvec\left(V_{1}\right)^{\top},\cdots,\tvec\left(V_{k}\right)^{\top}\right]^{\top}$.
Let $R_{i}\in\left\{ 0,1\right\} ^{\stackrel[m=0]{k}{\sum}L_{m}\times\stackrel[m=0]{k}{\sum}L_{m}}$
denote the $i^{\text{th}}$ instance of the randomization matrix,
for all $i\in\mathcal{I}\tq\left\{ 1,\cdots,J\right\} $, where $\mathcal{I}$
denotes the set of all possible switching instances. 

After every user-selected constant time period of $\delta t\in\RR_{>0}$
seconds, the randomization matrix $R_{i}$ switches to $R_{i+1}$,
where $R_{i+1}$ is randomly selected from all possible permutations
of randomization matrices. Each permutation is defined as {\footnotesize{}
\[
R_{i}\tq\left[\begin{array}{cccc}
R_{i,0} & \mathbf{0}_{L_{0}\times L_{1}} & \cdots & \mathbf{0}_{L_{0}\times L_{k}}\\
\mathbf{0}_{L_{1}\times L_{0}} & R_{i,1} & \cdots & \mathbf{0}_{L_{1}\times L_{k}}\\
\vdots & \vdots & \ddots & \vdots\\
\mathbf{0}_{L_{k}\times L_{0}} & \mathbf{0}_{L_{k}\times L_{1}} & \cdots & R_{i,k}
\end{array}\right],\quad\forall i\in\mathcal{I}.
\]
}For $i\in\left\{ 1,\cdots,J\right\} $ and $j\in\left\{ 1,\cdots,k\right\} $,
$R_{i,j}\in\left\{ 0,1\right\} ^{L_{j}\times L_{j}}$ is designed
to be a diagonal matrix, and the matrix $R_{i,0}$ is an identity
matrix. The number of ones on the diagonal of each $R_{i,j}$ is equal
to a user-selected constant number $n_{j}$, and the placement of
non-zero elements on the diagonal of the matrix $R_{i,j}$ randomly
changes after every $\delta t$ seconds.\footnote{Once the system reaches the steady state, the randomization can be
stopped in the sense that $R_{i}$ is replaced with identity matrices.
This can be considered as the final permutation of $R_{i}$ for all
$i\in\mathcal{I}$.} To illustrate the design of the randomization matrix $R_{i,j}$ and
the effect of dropout on the DNN architecture, the following example
is provided. 
\begin{example}
Consider $R_{i,2}\in\left\{ 0,1\right\} ^{3\times3}$, $V_{2}\in\RR^{3\times2}$,
and let $n_{2}=1$. Therefore, every $\delta t$ seconds, $n_{2}=1$
of the elements on the diagonal of $R_{i,2}$ are randomly set to
1 and the others are zeroed. The considered permutations of $R_{i,2}$
for $i\in\left\{ 1,2,3\right\} $ are {\tiny{}
\begin{gather*}
R_{1,2}=\left[\begin{array}{ccc}
1 & 0 & 0\\
0 & 0 & 0\\
0 & 0 & 0
\end{array}\right],\,R_{2,2}=\left[\begin{array}{ccc}
0 & 0 & 0\\
0 & 1 & 0\\
0 & 0 & 0
\end{array}\right],\,R_{3,2}=\left[\begin{array}{ccc}
0 & 0 & 0\\
0 & 0 & 0\\
0 & 0 & 1
\end{array}\right].
\end{gather*}
}Let $v_{p\times q}\in\RR$ denote to each individual weight of the
weight matrix $V_{2}$, for rows $p\in\left\{ 1,2,3\right\} $ and
columns $q\in\left\{ 1,2\right\} $. For $p\in\left\{ 1,2,3\right\} $,
let $\phi_{2,p}:\RR^{L_{2}}\to\RR^{L_{2}}$ denote to the activation
functions of the second layer such that the activation vector is $\phi_{2}=\left[\phi_{2,1},\,\phi_{2,2},\,\phi_{2,3}\right]^{\top}$.
Therefore, in the presence and absence of dropout matrix $\left(R_{3,2}V_{2}\right)^{\top}\phi_{2}$
and $V_{2}^{\top}\phi_{2}$ are respectively obtained as{\footnotesize{}
\begin{align}
\left(R_{3,2}V_{2}\right)^{\top}\phi_{2} & =\left[\begin{array}{cc}
v_{1,1} & v_{1,2}\\
v_{2,1} & v_{2,2}\\
v_{3,1} & v_{3,2}
\end{array}\right]^{\top}\left[\begin{array}{ccc}
0 & 0 & 0\\
0 & 0 & 0\\
0 & 0 & 1
\end{array}\right]^{\top}\left[\begin{array}{c}
\phi_{2,1}\\
\phi_{2,2}\\
\phi_{2,3}
\end{array}\right]\nonumber \\
 & =\left[\begin{array}{c}
v_{3,1}\phi_{2,3}\\
v_{3,2}\phi_{2,3}
\end{array}\right].\label{eq:eq}
\end{align}
\begin{align}
V_{2}^{\top}\phi_{2} & =\left[\begin{array}{cc}
v_{1,1} & v_{1,2}\\
v_{2,1} & v_{2,2}\\
v_{3,1} & v_{3,2}
\end{array}\right]^{\top}\left[\begin{array}{c}
\phi_{2,1}\\
\phi_{2,2}\\
\phi_{2,3}
\end{array}\right]\nonumber \\
 & =\left[\begin{array}{c}
v_{1,1}\phi_{2,1}+v_{2,1}\phi_{2,2}+v_{3,1}\phi_{2,3}\\
v_{1,2}\phi_{2,1}+v_{2,2}\phi_{2,2}+v_{3,2}\phi_{2,3}
\end{array}\right].\label{eq:eq2}
\end{align}
}Comparing (\ref{eq:eq}) and (\ref{eq:eq2}) suggests how the dropout
method deactivates the activation functions associated with the zeros
on the diagonal of the randomization matrix. Since the dropout matrix
is randomly generated, a new batch of weights are selected every $\delta t$
seconds.
\end{example}
The universal function approximation property states that the function
space of (\ref{eq: DNN}) is dense in $\mathcal{C}\left(\mathcal{X}\right)$,
where $\mathcal{C}\left(\mathcal{X}\right)$ denotes the space of
continuous functions over the compact set $\mathcal{X}\subseteq\RR^{n}$,
where $x\in\mathcal{X}$ \cite[Theorem 1.1]{Kidger.Lyons2020}. Therefore,
for all $j\in\left\{ 0,\cdots,k\right\} $, there exists a corresponding
vector of ideal weights $\theta^{*}\in\RR^{\stackrel[j=0]{k}{\sum}L_{j}L_{j+1}}$
such that $\sup_{x_{d}\in\Omega}\left\Vert f\left(x\right)-\Phi\left(x,R_{i},\theta^{*}\right)\right\Vert \leq\overline{\varepsilon}$.
Thus, the drift vector field can be modeled as
\begin{eqnarray}
f\left(x\right) & = & \Phi\left(x,R_{i},\theta^{*}\right)+\varepsilon\left(x\right),\label{eq: f(xd)}
\end{eqnarray}
where $\varepsilon:\RR^{n}\to\RR^{n}$ denotes an unknown function
reconstruction error that can be bounded as $\sup\left\Vert \varepsilon\left(x\right)\right\Vert \leq\overline{\varepsilon}$.
\begin{assumption}
\label{thm: bound on theta*}The vector of ideal weights can be bounded
by a known constant $\overline{\theta}\in\RR_{>0}$ as $\left\Vert \theta^{*}\right\Vert \leq\overline{\theta}$,
\cite[Assumption 1]{Lewis.Yesildirek.ea1996}.
\end{assumption}

\subsection{Adaptation Law}

To fulfill the tracking objective, the DNN model in (\ref{eq: f(xd)})
is used to estimate the unknown drift dynamics in (\ref{eq: dynamics}).
Since the ideal weights of the modeled DNN are unknown, adaptive estimates
of the weight matrices are developed to learn the unknown drift dynamics
$f\left(x\right)$. Let $\hat{\theta}:\RR_{\geq0}\to\RR^{\stackrel[j=0]{k}{\sum}L_{j}L_{j+1}}$
be defined as $\hat{\theta}\left(t\right)\tq\left[\tvec\left(\widehat{V}_{0}\right)^{\top},\tvec\left(\widehat{V}_{1}\right)^{\top},\cdots,\tvec\left(\widehat{V}_{k}\right)^{\top}\right]^{\top}$,
where $\widehat{V}_{j}:\RR^{L_{j}\times L_{j+1}}$, for all $j\in\left\{ 0,\cdots,k\right\} $,
denote the weight estimates. The corresponding weight estimation error
$\tilde{\theta}:\RR_{\geq0}\to\RR^{\stackrel[j=0]{k}{\sum}L_{j}L_{j+1}}$
is defined as $\tilde{\theta}\left(t\right)\tq\theta^{*}-\hat{\theta}\left(t\right)$.
Using the weight estimates $\hat{\theta}$, the adaptive estimate
of $f\left(x\right)$ can be represented as $\Phi\left(x,R_{i},\hat{\theta}\right)\tq\left(R_{i,k}\widehat{V}_{k}\right)^{\top}\phi_{k}\circ\cdots\circ\left(R_{i,1}\widehat{V}_{1}\right)^{\top}\phi_{1}\circ\left(R_{i,0}\widehat{V}_{0}\right)^{\top}x$.
The estimated DNN architecture can be written in a recursive relation
as
\begin{equation}
\widehat{\Phi}_{j}=\begin{cases}
\left(R_{i,j}\widehat{V}_{j}\right)^{\top}\phi_{j}\left(\widehat{\Phi}_{j-1}\right), & j=\left\{ 1,...,k\right\} ,\\
\left(R_{i,0}\widehat{V}_{0}\right)^{\top}x, & j=0,
\end{cases}\label{eq: DNN - recursive}
\end{equation}
where $\widehat{\Phi}_{j}$ is the shorthand notation for $\widehat{\Phi}_{j}\tq\Phi_{j}\left(x,R_{i},\hat{\theta}\right)$,
and $\widehat{\Phi}=\widehat{\Phi}_{k}$. Based on the subsequent
stability analysis, the adaptation law for DNN weight estimates is
designed as
\begin{equation}
\dot{\hat{\theta}}\tq\text{proj}\left(\Gamma_{\theta}\widehat{\Phi}^{\prime\top}e\right),\label{eq: update law}
\end{equation}
where $\Gamma_{\theta}\in\RR^{\stackrel[j=0]{k}{\sum}L_{j}L_{j+1}\times\stackrel[j=0]{k}{\sum}L_{j}L_{j+1}}$
denotes a positive-definite adaptation gain matrix, and $\widehat{\Phi}^{\prime}$
is a shorthand notation for the Jacobian $\widehat{\Phi}^{\prime}\triangleq\frac{\partial\Phi\left(x,R_{i},\hat{\theta}\right)}{\partial\hat{\theta}}$.
The Jacobian $\widehat{\Phi}^{\prime}$ can be represented as $\widehat{\Phi}^{\prime}\triangleq\left[\widehat{\Phi}_{0}^{\prime},...,\widehat{\Phi}_{k}^{\prime}\right],$
where the shorthand notation $\widehat{\Phi}_{j}^{\prime}$ is defined
as $\widehat{\Phi}_{j}^{\prime}\triangleq\frac{\partial\Phi_{j}\left(x,R_{i},\hat{\theta}\right)}{\partial\hat{\theta}}$,
for all $j\in\left\{ 0,...,k\right\} $. The projection operator is
incorporated in the update law to ensure that $\hat{\theta}\left(t\right)\in\Lambda\tq\bigg\{\theta\in\RR^{\stackrel[j=0]{k}{\sum}L_{j}L_{j+1}}:\left\Vert \theta\right\Vert \leq\overline{\theta}\bigg\}$,
for all $t\geq0$. Since $\left\Vert \theta^{*}\right\Vert \leq\overline{\theta}$
and $\left\Vert \hat{\theta}\right\Vert \leq\overline{\theta}$, using
the definition of $\tilde{\theta}$, it can be shown that $\left\Vert \tilde{\theta}\right\Vert \leq2\overline{\theta}$.
Using (\ref{eq: DNN - recursive}), the chain rule, and the properties
of the vectorization operator, the Jacobians $\widehat{\Phi}_{0}^{\prime}$
and $\widehat{\Phi}_{j}^{\prime}$, for all $j=\left\{ 1,\cdots,k\right\} $
and $i\in\mathcal{I}$, can respectively be calculated as {\footnotesize{}
\begin{align}
\widehat{\Phi}_{0}^{\prime} & \tq\left(\stackrel[l=1]{\overset{\curvearrowleft}{k}}{\prod}\left(R_{i,l}\widehat{V}_{l}\right)^{\top}\widehat{\phi}_{l}^{\prime}\right)\left(\left(x^{\top}R_{i,0}\right)\otimes I_{L_{1}}\right),\nonumber \\
\widehat{\Phi}_{j}^{\prime} & \tq\left(\stackrel[l=j+1]{\overset{\curvearrowleft}{k}}{\prod}\left(R_{i,l}\widehat{V}_{l}\right)^{\top}\widehat{\phi}_{l}^{\prime}\right)\left(\left(\widehat{\phi}_{j}^{\top}R_{i,j}\right)\otimes I_{L_{j+1}}\right),\label{eq: phi_prime_hat}
\end{align}
}where $\hat{\phi}_{j}$ and the Jacobian $\hat{\phi}_{j}^{\prime}$
are the short-hand notations for $\hat{\phi}_{j}\tq\phi_{j}\left(\widehat{\Phi}_{j-1}\right)$
and $\hat{\phi}_{j}^{\prime}\tq\phi_{j}^{\prime}\left(\widehat{\Phi}_{j-1}\right)=\frac{\partial\widehat{\Phi}_{j}}{\partial\hat{\theta}}$,
respectively.
\begin{rem}
\label{rem:co-adaptation}The presence of matrix $R_{i}$ for all
$i\in\mathcal{I}$ in (\ref{eq: phi_prime_hat}) mitigates co-adaptation
by reducing the interdependency of weights in the adaptation law.
\end{rem}

\subsection{Closed-Loop Error System}

The designed DNN estimate is used in the developed controller to approximate
the unknown drift vector field in (\ref{eq: dynamics}). By incorporating
the developed adaptive DNN estimate, the controller in (\ref{eq: controller})
is designed such that the state $x$ tracks the desired trajectory
$x_{d}$ despite inactivation of weights associated with a randomly
selected batch of neurons. Based on the subsequent stability analysis,
the control input is designed as
\begin{equation}
u\left(t\right)\tq\dot{x}_{d}-\widehat{\Phi}-k_{e}e-k_{s}\text{sgn}\left(e\right),\label{eq: controller}
\end{equation}
where $k_{e},k_{s}\in\RR_{>0}$ are constant control gains. Taking
the time-derivative of (\ref{eq: error}) and substituting (\ref{eq: dynamics}),
(\ref{eq: f(xd)}), and the designed controller in (\ref{eq: controller})
and canceling cross-terms yields the closed-loop error system as{\small{}
\begin{alignat}{1}
\dot{e}\left(t\right)=\Phi\left(x,R_{i},\theta^{*}\right)-\widehat{\Phi}+\varepsilon\left(x\right)-k_{s}\text{sgn}\left(e\right)-k_{e}e.\label{eq: pre-CLES}
\end{alignat}
}To address the technical challenges in deriving adaptation the DNN
weights, many results use Taylor series approximation based techniques
\cite[Eq. 22]{Lewis.Yesildirek.ea1996,Patil.Le.ea.2022,Patil.Le.ea2022,Griffis.Patil.ea23_2}.
Applying a first-order Taylor series approximation-based error model
on $\Phi\left(x,R_{i},\theta^{*}\right)-\widehat{\Phi}$ yields
\begin{alignat}{1}
\Phi\left(x,R_{i},\theta^{*}\right)-\widehat{\Phi} & =\widehat{\Phi}^{\prime}\tilde{\theta}+\mathcal{O}^{2}\left(\left\Vert \tilde{\theta}\right\Vert \right),\label{eq: TSA}
\end{alignat}
where $\mathcal{O}\left(\left\Vert \tilde{\theta}\right\Vert \right)$
denotes the higher-order terms that can be bounded as $\left\Vert \mathcal{O}^{2}\left(\left\Vert \tilde{\theta}\right\Vert \right)\right\Vert \leq\Delta$,
where $\Delta\in\RR_{>0}$ denotes a known constant \cite[Eq. 18]{Patil.Le.ea2022}.
Substituting (\ref{eq: TSA}) into (\ref{eq: pre-CLES}) yields
\begin{alignat}{1}
\dot{e}=\widehat{\Phi}^{\prime}\tilde{\theta}+\mathcal{O}^{2}\left(\left\Vert \tilde{\theta}\right\Vert \right)+\varepsilon\left(x\right)-k_{s}\text{sgn}\left(e\right)-k_{e}e.\label{eq: CLES}
\end{alignat}
To facilitate the subsequent stability analysis, let $z:\RR_{\geq0}\to\RR^{\psi}$
denote the concatenated error system defined as 
\begin{equation}
z\left(t\right)\tq\left[e^{\top}\left(t\right),\,\tilde{\theta}^{\top}\left(t\right)\right]^{\top},\label{eq: z}
\end{equation}
where $\psi\tq n+\stackrel[j=0]{k}{\sum}L_{j}L_{j+1}$. Additionally,
let $\dot{z}=h\left(z,t\right)$, where $h:\RR^{\psi}\times\RR_{\geq0}\to\RR^{\psi}$
is as
\begin{multline}
h\left(z,t\right)\tq\left[\begin{array}{c}
\left(\begin{array}{c}
\widehat{\Phi}^{\prime}\tilde{\theta}+\mathcal{O}^{2}\left(\left\Vert \tilde{\theta}\right\Vert \right)\\
+\varepsilon\left(x\right)-k_{s}\text{sgn}\left(e\right)-k_{e}e
\end{array}\right)\\
-\text{proj}\left(\Gamma_{\theta}\widehat{\Phi}^{\prime\top}e\right)
\end{array}\right].\label{eq: h(z,t)}
\end{multline}

\section{Stability Analysis}

Let $V_{L}:\RR^{\psi}\to\RR_{\geq0}$ denote the Lyapunov function
candidate defined as 
\begin{equation}
V_{L}\left(z\right)\tq\frac{1}{2}e^{\top}e+\frac{1}{2}\tilde{\theta}^{\top}\Gamma_{\theta}^{-1}\tilde{\theta}.\label{eq: Lyap. func. candidate}
\end{equation}
Given the known constants $\underline{\alpha},\overline{\alpha}\in\RR_{>0}$,
the Lyapunov function candidate satisfies the following inequality:
\begin{equation}
\underline{\alpha}\nzz\leq V_{L}\left(z\right)\leq\overline{\alpha}\nzz.\label{eq: bounds on V_L}
\end{equation}
Let the open and connected sets $\mathcal{B}\in\RR^{\psi}$ and $\Upsilon\subseteq\mathcal{X}$
be defined as $\mathcal{B}\tq\left\{ \varsigma\in\RR^{\psi}:\lVert\varsigma\rVert\leq\omega\sqrt{\nicefrac{\underline{\alpha}}{\overline{\alpha}}}\right\} $
and $\Upsilon\tq\left\{ \varsigma\in\mathcal{X}:\lVert\varsigma\rVert<\overline{x_{d}}+\omega\right\} $.
Theorem \ref{thm: Stability } uses the non-smooth analysis technique
in \cite{Fischer.Kamalapurkar.ea2013} to establish the invariance
properties of Fillipov solutions to $\dot{z}$ and to guarantee asymptotic
convergence of the tracking error, $e$.
\begin{thm}
\label{thm: Stability } The controller designed in (\ref{eq: controller})
and the DNN update law developed in (\ref{eq: update law}) guarantee
asymptotic tracking error convergence for the dynamical system in
(\ref{eq: dynamics}) in the sense that $\underset{t\to\infty}{\lim}\left\Vert e\left(t\right)\right\Vert =0$,
given $\left\Vert z\left(t_{0}\right)\right\Vert \in\mathcal{B}$
and that the gain condition $k_{s}>\overline{\varepsilon}+\Delta$
is satisfied.
\end{thm}
\begin{IEEEproof}
Let $\partial V_{L}$ denote the Clarke gradient of $V_{L}$ defined
in \cite[p. 39]{Clarke1990}. Since the Lyapunov function candidate
is continuously differentiable, $\partial V_{L}(z)=\{\nabla V_{L}(z)\}$,
where $\nabla$ denotes the standard gradient operator. From (\ref{eq: h(z,t)}),
it can be concluded that for all $i\in\mathcal{I}$, $V_{L}$ satisfies
the following differential equation
\begin{align}
\dot{V}_{L} & \overset{\text{a.a.t.}}{\in}\overset{\sigma\in\partial V_{L}\left(z\right)}{\bigcap}\sigma^{\top}\text{K}\left[h\right]\left(z,t\right)\nonumber \\
 & =\nabla V_{L}^{\top}\left(z\right)\text{K}\left[h\right]\left(z,t\right)\nonumber \\
 & =e^{\top}\bigg(\widehat{\Phi}^{\prime}\tilde{\theta}+\mathcal{O}^{2}\left(\left\Vert \tilde{\theta}\right\Vert \right)+\varepsilon\left(x\right)-k_{e}e\nonumber \\
 & \quad-k_{s}\text{K}\left[\text{sgn}\right]\left(e\right)\bigg)-\tilde{\theta}^{\top}\Gamma_{\theta}^{-1}\text{K}\left[\text{proj}\right]\big(\Gamma_{\theta}\widehat{\Phi}^{\prime\top}e\big).\label{eq: Vdot}
\end{align}
Using \cite[Lemma E.1.IV]{Krstic1995}, the bounds on $\mathcal{O}^{2}\left(\left\Vert \tilde{\theta}\right\Vert \right)$
and $\varepsilon\left(x\right)$, the fact that $\text{K}\left[\text{proj}\right]\left(\cdot\right)$
is the set of convex combinations of $\text{proj}\left(\cdot\right)$
and $\left(\cdot\right)$, and therefore, $-\tilde{\theta}^{\top}\Gamma_{\theta}^{-1}\text{K}\left[\text{proj}\right]\left(\Gamma_{\theta}\widehat{\Phi}^{\prime\top}e\right)\leq-\tilde{\theta}^{\top}\widehat{\Phi}^{\prime\top}e$,
and canceling cross terms, (\ref{eq: Vdot}) can be upper-bounded
as
\begin{align*}
\dot{V}_{L} & \overset{\text{a.a.t.}}{\leq}-k_{e}\nee-k_{s}\ne+\ne\left(\Delta+\overline{\varepsilon}\right),\quad\forall i\in\mathcal{I}.
\end{align*}
Selecting the gain $k_{s}$ according to the gain condition in Theorem
\ref{thm: Stability } yields
\begin{equation}
\dot{V}_{L}\overset{\text{a.a.t.}}{\leq}-k_{e}\nee.\label{eq: bounds dot(V_L)}
\end{equation}
From the inequality obtained in (\ref{eq: bounds dot(V_L)}), \cite[Corollary 1]{Fischer.Kamalapurkar.ea2013}
can be invoked to conclude that $z\in\Linf$ and $\underset{t\to\infty}{\lim}\left\Vert e\left(t\right)\right\Vert =0$.
Due to the facts that $\widehat{\Phi}$ is smooth for all $i\in\mathcal{I}$,
$x\in\Omega$, and $\left\Vert \hat{\theta}\right\Vert \leq\overline{\theta}$,
$\widehat{\Phi}\in\Linf$. Since $\dot{x}_{d}\in\Linf$, $e\in\Linf$,
and $\widehat{\Phi}\in\Linf$, $u\in\Linf$. To show that $x\in\mathcal{X}$,
and therefore the universal function approximation property holds,
let $\left\Vert z\left(t_{0}\right)\right\Vert \in\mathcal{B}$. Since
$\lVert z\left(t_{0}\right)\rVert\leq\omega\sqrt{\nicefrac{\underline{\alpha}}{\overline{\alpha}}}$,
using (\ref{eq: bounds on V_L}), $\lVert e\left(t\right)\rVert\leq\omega$.
Hence, using (\ref{eq: error}), $\left\Vert x\right\Vert $ can be
bounded as $\left\Vert x\right\Vert \leq\overline{x_{d}}+\omega$.
Therefore, if $z\left(t_{0}\right)\in\mathcal{B}$, then $x\in\Upsilon\subseteq\mathcal{X}$.
\end{IEEEproof}

\section{Simulation}

To demonstrate the efficacy of the Lb-DDNN adaptive controller, simulations
are performed on a three-dimensional nonlinear system, and $f$ in
(\ref{eq: dynamics}) is modeled as
\[
f=\left[\begin{array}{c}
x_{1}x_{2}^{2}\tanh\left(x_{2}\right)+\sin\left(x_{1}\right)^{2}\\
\cos\left(x_{1}+x_{2}+x_{3}\right)^{3}-\exp\left(x_{2}\right)^{2}+x_{1}x_{2}\\
x_{3}^{2}\log\left(1+\text{abs}\left(x_{1}-x_{2}\right)\right)
\end{array}\right],
\]
where $x\tq\left[x_{1},x_{2},x_{3}\right]^{\top}:\RR_{\geq0}\to\RR^{3}$
denotes the system state. Three simulation experiments are performed
for $10\,\text{sec}$ with initial condition $x\left(0\right)=\left[5,1,-5\right]^{\top}$.
The desired trajectory is selected as $x_{d}\left(t\right)=\left[\sin\left(2t\right),-\cos\left(t\right),\sin\left(3t\right)+\cos\left(-2t\right)\right]^{\top}$.
The DNN used in the simulations has $k=7$ inner layers with $L=10$
neurons in each hidden layer and contained hyperbolic tangent activation
functions. The first set of simulations are performed to compare the
baseline DNN-based adaptive controller in \cite{Patil.Le.ea.2022}
and the Lb-DDNN adaptive controller in (\ref{eq: update law}) and
(\ref{eq: controller}). The second set of simulations are performed
to examine the effect of $\delta t$ on the performance of the propose
method. The third set of simulations are performed to compare the
performance in the absence and presence of dropout deactivation after
the transient period. In all simulations, the DNN weight estimates
are initialized randomly from the normal distribution $\mathcal{N}\left(0,10\right)$.
The control gains in (\ref{eq: controller}) are selected as $k_{e}=10.5$
and $k_{s}=1.5$. The learning gain for the baseline DNN is selected
as $\Gamma_{\theta}=100I_{670\times670}$. In the first two sets of
simulations, the randomization is activated for the first $2\,\text{sec}$
where the system is in the transient stage. After $2\,\text{sec}$,
all the randomization matrices change to identity matrices. For the
first $2\,\text{sec}$, the learning gain of the dropout DNN update
law in (\ref{eq: update law}) is selected as $\Gamma_{\theta}=100I_{670\times670}$,
and after $2\,\text{sec}$, the learning gain changes to $\Gamma_{\theta}=40I_{670\times670}$.
In the transient stage of the first and third set of simulations,
the matrices $R_{i}$ change to $R_{i+1}$ every $\delta t=0.1\,\text{sec}$. 

\begin{figure}
\begin{centering}
\includegraphics[viewport=30bp 30bp 900bp 900bp,scale=0.27]{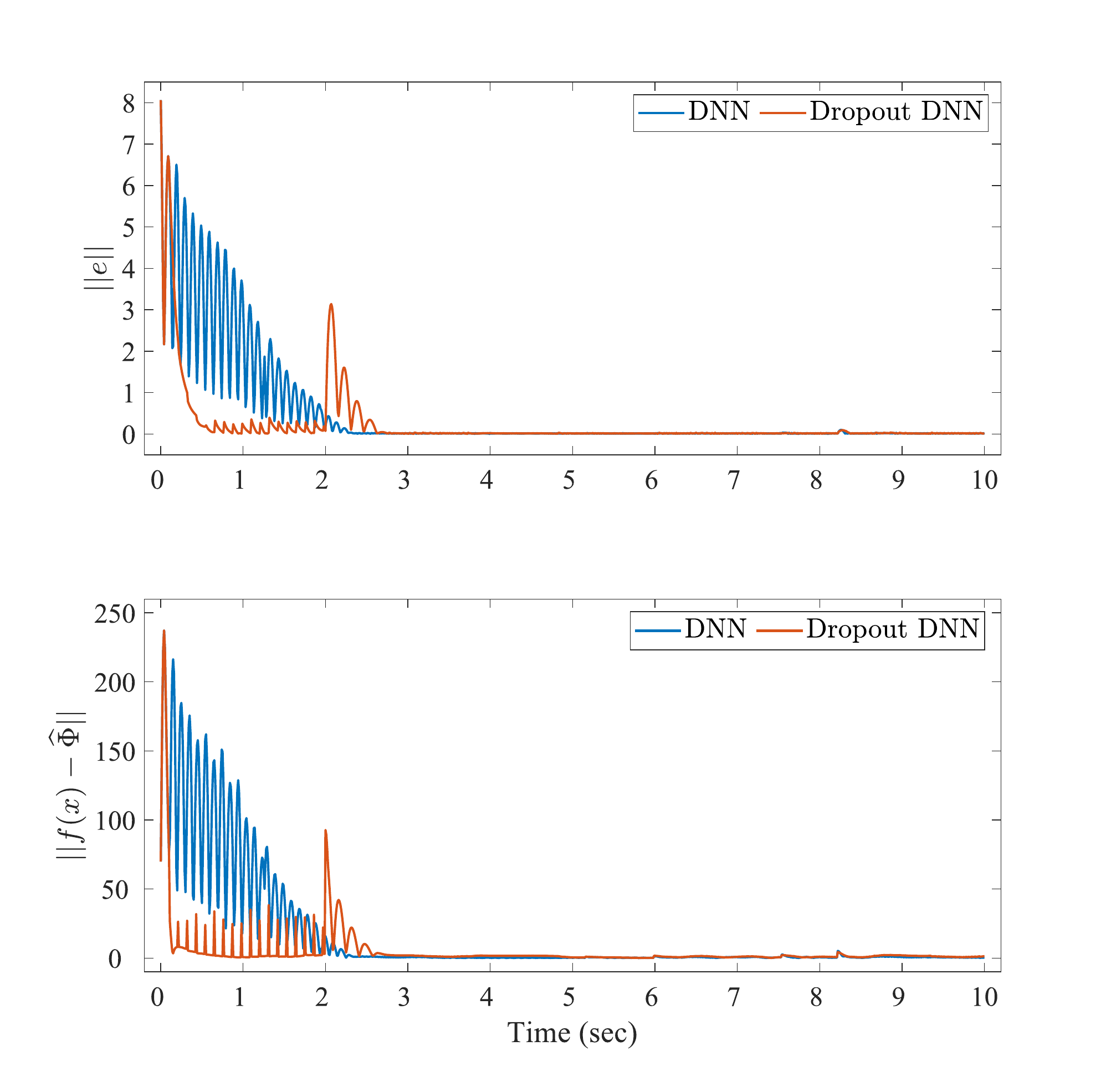}
\par\end{centering}
\centering{}\caption{Performance of the tracking error (top) and function approximation
error (bottom) over time comparing the Lb-DDNN controller and the
adaptive DNN controller developed in \cite{Patil.Le.ea.2022}\label{fig: Sim result}.}
\end{figure}

The performance results of the simulations are presented in Table
\ref{tab:sim result}. As shown in the first subplot of Figure \ref{fig: Sim result},
the tracking error for the dropout DNN converges significantly faster
than the baseline DNN. Specifically, the dropout DNN results in convergence
to the final error after approximately $0.5\,\text{sec}$, roughly
four times faster than that of the baseline controller. Despite the
jump in the tracking error after the deactivation of the dropout of
the DNN weights, the dropout DNN still yields the norm of the root
mean square tracking error of $0.81$, which shows a $38.32\%$ improvement
when compared to the baseline DNN adaptive controller. Moreover, the
baseline DNN controller presents more oscillatory behavior within
the transient period than the dropout DNN controller. The oscillatory
behavior in the baseline DNN is due to interdependency of weights
in the adaptation. However, as stated in Remark \ref{rem:co-adaptation},
the dropout DNN mitigates co-adaptation in the adaptation law, thus
yielding less oscillatory behavior. As shown in the second subplot
of Figure \ref{fig: Sim result}, the function approximation error
for the dropout DNN controller rapidly converges after less than $0.2\,\text{sec}$
but takes approximately $2\,\text{sec}$ to converge with the baseline
DNN controller. Although there is a jump in the function approximation
error after the deactivation of the DNN, the dropout DNN controller
demonstrated a $53.67\%$ improvement in function approximation with
the norm of the root mean square function approximation error of $19.44$.
Thus, the developed dropout adaptive DNN architecture resulted in
better transient behavior and improved tracking and function approximation
performance with a $50.44\%$ lower control effort when compared to
the baseline adaptive DNN controller developed in \cite{Patil.Le.ea.2022}. 

\begin{figure}
\begin{centering}
\includegraphics[viewport=30bp 30bp 900bp 900bp,clip,scale=0.27]{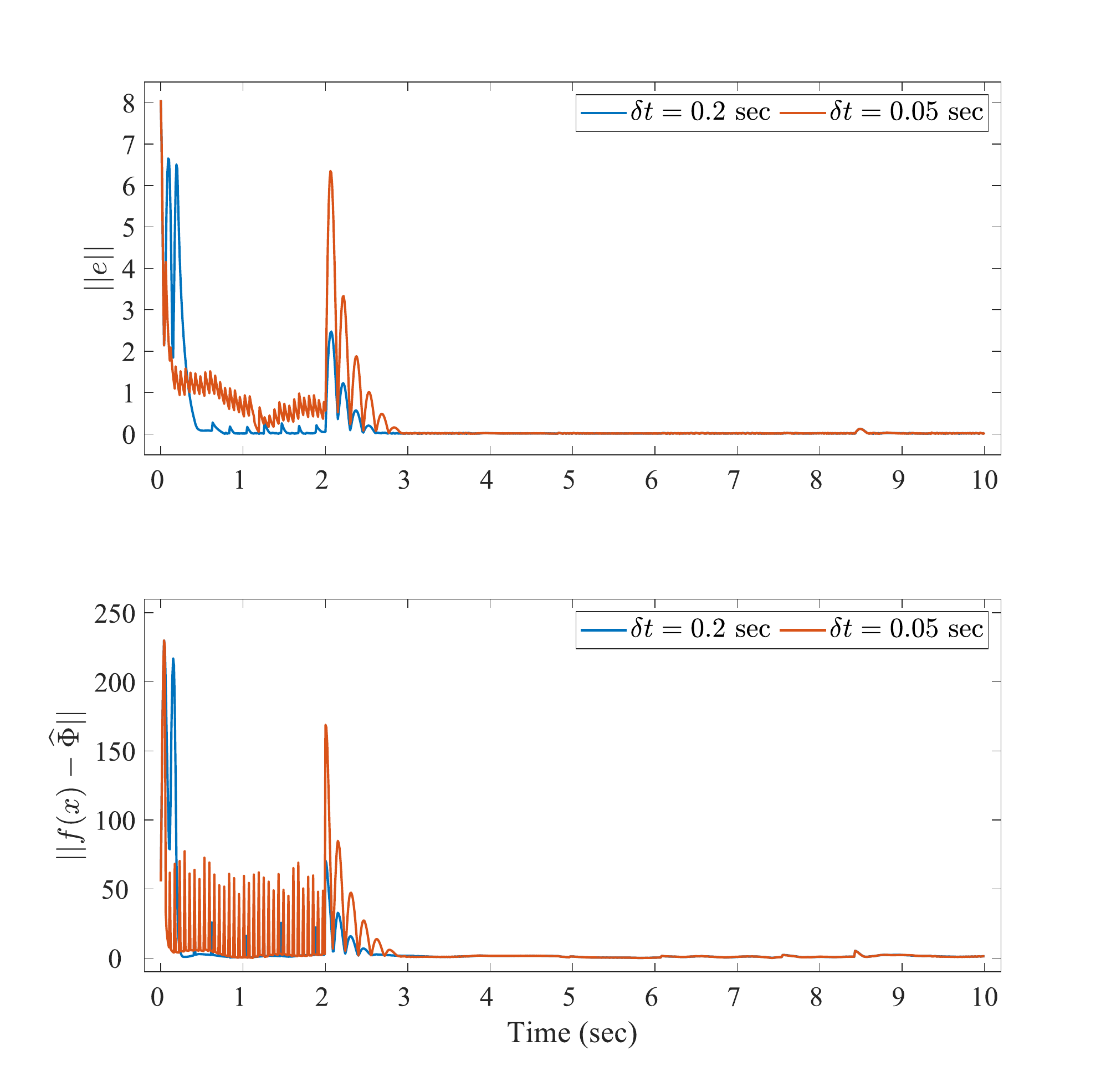}
\par\end{centering}
\centering{}\caption{Performance of the tracking error (top) and function approximation
error (bottom) over time for dropout DNN with two different $\delta t$
values.\label{fig: Sim result-1}}
\end{figure}

To examine the effect of selecting different $\delta t$, simulations
are performed with $\delta t=0.2\,\text{sec}$ and $\delta t=0.05\,\text{sec}$
using the Lb-DDNN controller. As shown in Figure \ref{fig: Sim result-1},
reducing $\delta t$ causes more spikes in the tracking and function
approximation performances. Although the differences between the the
tracking and function approximation errors are not significant, reducing
$\delta t$ is found to cause more spikes in the plots, as shown in
Figure \ref{fig: Sim result-1}.

\begin{figure}
\begin{centering}
\includegraphics[viewport=30bp 30bp 900bp 900bp,scale=0.27]{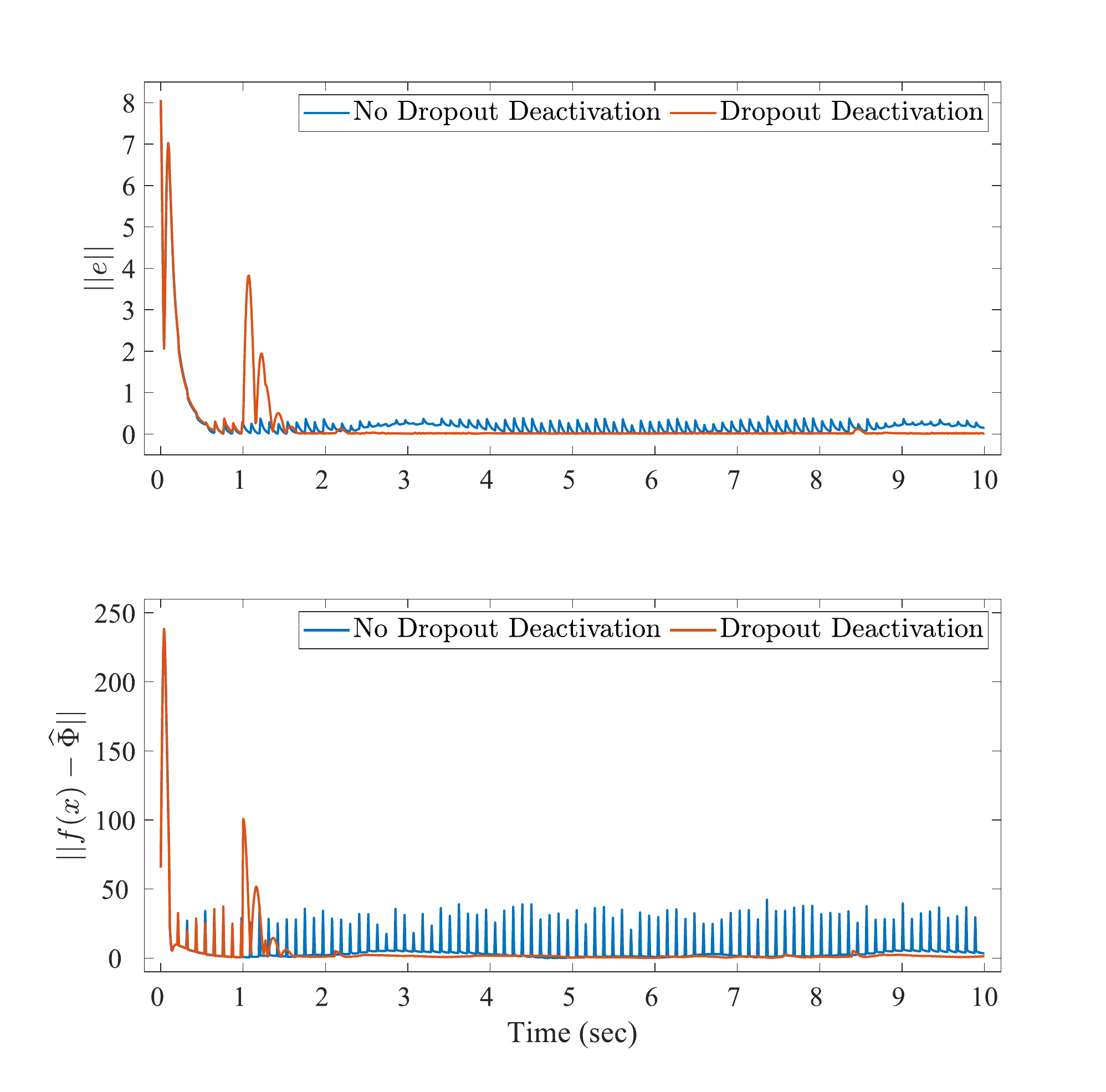}
\par\end{centering}
\centering{}\caption{Performance of the tracking error (top) and function approximation
error (bottom) over time in the presence and absence of dropout deactivation
in the steady state (after $1\,\text{sec}$).\label{fig: Sim result-2}}
\end{figure}

The third set of simulations examine the performance of the developed
dropout DNN controller under two cases; dropping out the neurons for
the entire duration of the simulation, and deactivating dropout after
$1\,\text{sec}$. As shown in Figure \ref{fig: Sim result-2}, for
both cases, the difference between the tracking and function approximation
performances are insignificant during the first second, as expected.
Once the dropout is deactivated after $1\,\text{sec}$, there is an
overshoot in both tracking and function approximation errors which
does not occur when dropout is maintained throughout the simulation
duration. However, not deactivating the dropout after the transient
period leads to more spikes in both tracking and function approximation
error in the steady state stage. Despite the increase in the the tracking
error, deactivation of the dropout in the steady state leads to lower
control input and function approximation error as shown in Table \ref{tab:sim result}.

\begin{table}
\caption{Performance comparison results\label{tab:sim result}}

\centering{}{\scriptsize{}}%
\begin{tabular}{|c|c|c|c|}
\hline 
{\scriptsize{}DNN Architecture} & {\scriptsize{}$\ne$} & {\scriptsize{}$\left\Vert f-\widehat{\Phi}\right\Vert $} & {\scriptsize{}$\left\Vert u\right\Vert $}\tabularnewline
\hline 
{\scriptsize{}DNN} & {\scriptsize{}$1.32$} & {\scriptsize{}$41.81$} & {\scriptsize{}$780.19$}\tabularnewline
\hline 
{\scriptsize{}Dropout DNN} & {\scriptsize{}$0.81$} & {\scriptsize{}$19.44$} & {\scriptsize{}$386.67$}\tabularnewline
\hline 
{\scriptsize{}Dropout DNN, $\delta t=0.2\,\text{sec}$} & {\scriptsize{}$0.88$} & {\scriptsize{}$23.87$} & {\scriptsize{}$461.83$}\tabularnewline
\hline 
{\scriptsize{}Dropout DNN, $\delta t=0.05\,\text{sec}$} & {\scriptsize{}$0.90$} & {\scriptsize{}$22.64$} & {\scriptsize{}$430.79$}\tabularnewline
\hline 
{\scriptsize{}Dropout DNN,} & \multirow{2}{*}{{\scriptsize{}$0.78$}} & \multirow{2}{*}{{\scriptsize{}$20.40$}} & \multirow{2}{*}{{\scriptsize{}$399.93$}}\tabularnewline
{\scriptsize{}no dropout deactivation} &  &  & \tabularnewline
\hline 
{\scriptsize{}Dropout DNN, } & \multirow{2}{*}{{\scriptsize{}$0.86$}} & \multirow{2}{*}{{\scriptsize{}$20.00$}} & \multirow{2}{*}{{\scriptsize{}$387.32$}}\tabularnewline
{\scriptsize{}dropout deactivation after $1\,\text{sec}$} &  &  & \tabularnewline
\hline 
\end{tabular}
\end{table}

\section{Conclusion}

A dropout DNN-based adaptive controller is developed for general continuous
nonlinear systems. Leveraging the stability-derived DNN update law
in \cite{Patil.Le.ea.2022} and inspired by the dropout technique,
the developed dropout DNN controller improves function approximation
performance and yields faster learning when compared to the DNN controllers
without dropout. A Lyapunov-based stability analysis is performed
to guarantee stability in the sense that the tracking error asymptotically
converges to zero. Simulation results show $38.32\%$ and $53.67\%$
improvement in the tracking error and function approximation error,
respectively, with a $50.44\%$ reduced control effort when compared
to the baseline adaptive DNN controller. Additional simulations showed
the effect of dropout during both transient and steady state periods
and how modifying dropout parameters, i.e., $\delta t$, can effect
system performance. Using the established Lb-DDNN framework, future
work can explore implementation questions related to the dropout regularization
such as changes in $\delta t$, the number of neurons that are randomly
selected, and dropout deactivation strategies.

\bibliographystyle{ieeetr}
\bibliography{master,ncr,encr}

\end{document}